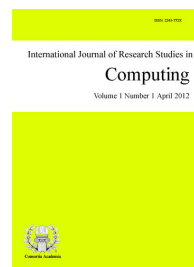

# On replacing PID controller with ANN controller for DC motor position control

Aamir, Muhammad ✉
*Shaheed Zulfikar Ali Bhutto Institute of Science and Technology, Karachi, Pakistan*
(*aamir.nbpit@yahoo.com*)

*Abstract*

The process industry implements many techniques with certain parameters in its operations to control the working of several actuators on field. Amongst these actuators, DC motor is a very common machine. The angular position of DC motor can be controlled to drive many processes such as the arm of a robot. The most famous and well known controller for such applications is PID controller. It uses proportional, integral and derivative functions to control the input signal before sending it to the plant unit. In this paper, another controller based on Artificial Neural Network (ANN) control is examined to replace the PID controller for controlling the angular position of a DC motor to drive a robot arm. Simulation is performed in MATLAB after training the neural network (supervised learning) and it is shown that results are acceptable and applicable in process industry for reference control applications. The paper also indicates that the ANN controller can be less complicated and less costly to implement in industrial control applications as compared to some other proposed schemes.

# On replacing PID controller with ANN controller for DC motor position control

1. **Introduction**

In order to control the parameters of industrial processes, there are various kinds of actuators on field. Actuator is a driver that runs some mechanical activity. For example, if a process needs to open a valve for fluid motion or move a robotic arm for some appropriate action, there will be a motor with specific applied controls such as the speed and angular position control. DC (Direct Current) motors are often used in various industrial applications where a wide range of responses are required to follow a predetermined trajectory of speed or position under variable load (Faramarzi & Sabahi, 2011). There are various types of control mechanisms that may be applied on the speed and angular position of a DC motor, depending upon the accuracy required.

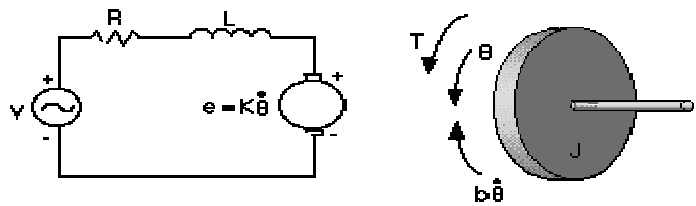

*Figure 1*. Electric circuit of a DC motor

In figure 1, electric circuit of a DC motor is shown that governs its rotation for desired velocity or position. The dynamics of a DC motor may be explained by the following equations:

$$v = Ri + L\frac{di}{dt} + e \qquad (1)$$

$$T = Ki \qquad (2)$$

In equation 1, *v* is voltage applied across armature, *R* is Armature Resistance, *i* is Armature Current, *L* is Armature Inductance and *e* is back electromotive force (emf) produced across the armature terminals upon its rotation. In equation 2, *T* is Torque, *K* is motor constant representing torque constant and back emf constant, *i* is Armature Current.

The DC motor's torque is also represented by the following relation:

$$T = J\frac{d^2\theta}{dt^2} + b\frac{d\theta}{dt} \qquad (3)$$

In equation 3, *T* is Torque, *J* is moment of inertia of motor and its load, *Θ* is angular displacement of motor's shaft and *b* is frictional constant of motor and its load (Thomas & Poongodi, 2009). In order to control the velocity or position of a DC motor, a torque is applied across its armature with controlled parameters. This torque is controlled by a calculated voltage signal at the input.

The most common control application for speed and position controls of DC motors with high accuracy in industry is PID (Proportional-Integral-Derivative) control. In this control, a special circuit or program is required to control the output of controller and prevent the plant from unwanted disturbances or instability in performance. The input is not directly fed into the plant or actuator, but it goes into the controller where the programmed logic manipulates the signal for desired response and finally the controlled signal is sent to the plant or actuator (DC motor in case of a robot arm).

Neural networks model human neural systems through computerized algorithm. They are capable of parallel computations and distributive storage of information like human brain. In recent years, they have been widely used for optimum calculations and processes in industrial controls, communications, chemistry and petroleum. They are on the rise for use in many highly sensitive control mechanisms such as flight controls and implementation of high security devices (Xu et al., 2012; Yerramalla & Cukic, 2005).

2. **Negative Feedback with PID Control**

DC motors are quite popular in process industry due to various control characteristics. However due to some disadvantages, the position control systems were implemented in process industry using stepper motors and induction motors. The operation of stepper motors is open-loop which ultimately produces very low performance results. The step response is very poor having significant overshoot and long settling time. Therefore, feedback control systems have been proposed for stepper motor position control systems. The same is also applicable on DC motor position control systems with acceptable results. Figure 2 shows the basic block diagram of a feedback control system.

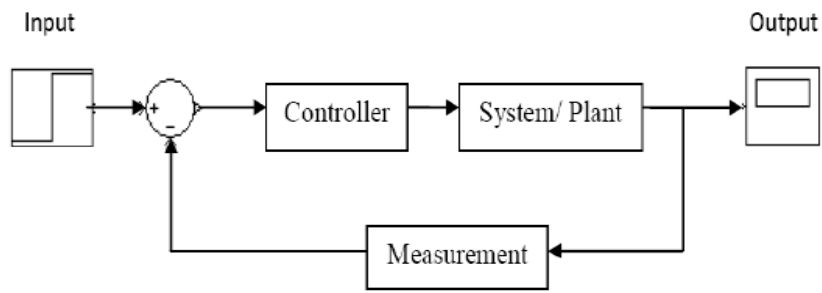

*Figure 2.* Basic block diagram of a feedback control system

In general, negative feedback is good but if the controller is incorrectly designed, the feedback system becomes unstable and oscillates. Therefore, it is very important to make sure that the system is stable before it goes live on the actual process.

PID control uses three mathematical control functions and applies them to input signals for desired outputs. Proportional value determines the response of the system towards current error. Integral value fastens the response introduced by proportional factor. However, increasing value of integral part makes the system to oscillate with overshoots. Derivative part reduces the overshoots introduced by integral part. However, increasing value of derivative control makes the response slow. In MATLAB, we analyze step response of our robot arm based on DC motor position control (the plant) by applying PID control with values of P=25, I=17 and D=11. These values were selected to obtain an appropriate result in the system under consideration.

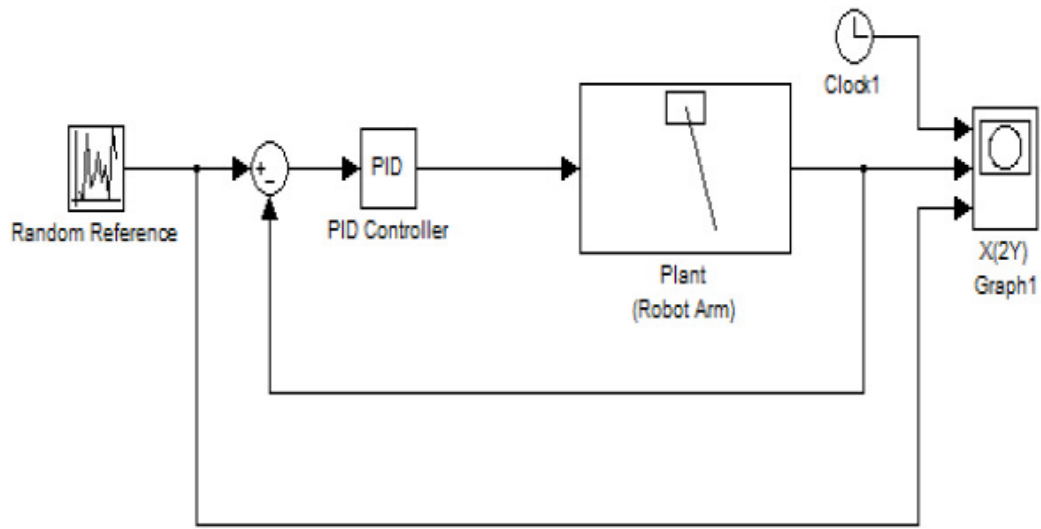

*Figure 3.* PID control for robot arm (DC motor position control) in MATLAB

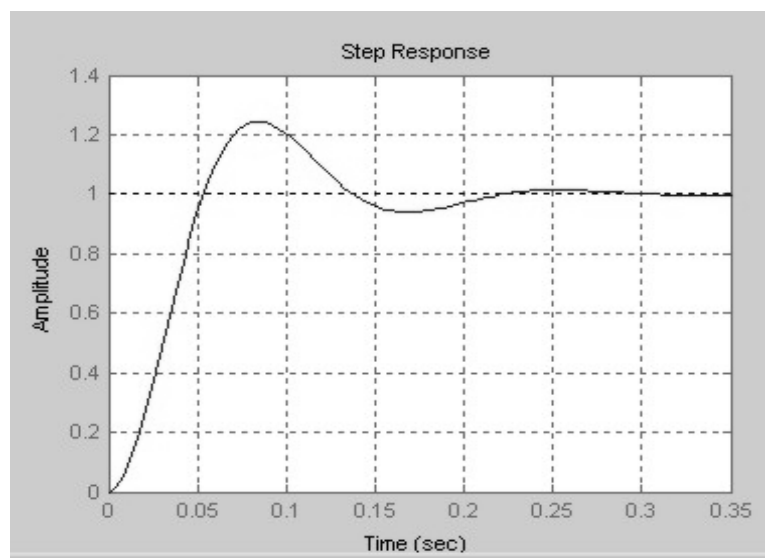

*Figure 4.* Step response of PID control for robot arm

The response shown in figure 4 has time value on X Axis and amplitude level on Y Axis. The curve represents the output in response to a given step input. The supplied values of P, I and D have produced a good output with less overshoot and acceptable steady state response. PID control is widely used in process industry due to its good characteristics and response dynamics. However, PID control has some limitations too. In environments where high accuracy is required, such as robotics and guided manipulations, the control factor demands special attention and accuracy for desired results. Many control implementations are composed of fixed (constant) gain controllers such as PI (Proportional-Integral) and PID (Proportional-Integral-Derivative). They usually stabilize linear systems and need accurate models describing dynamics of the system. Accurate models are hard to achieve as unknown dynamics cannot be modeled accurately such as environmental noise, saturation and parameter drifts. In some cases, sudden variations in load also make a system unstable if not properly controlled. It can also be a case in motors as load due to their nonlinear mechanical properties. Therefore, many research attempts have come up with adaptive control implementations such as sliding mode control, model reference adaptive control, fuzzy PID and neural network PID (Li & Zou, 2011; Kadam et al., 2010).

## 3. Applying Artificial Neural Network (ANN) Controller

Artificial Neural Networks are famous learning models for their ability to cope with the demands of a changing environment (Liu et al., 2007). In this study, we analyze the application of Artificial Neural Network (ANN) controller in process industry as a replacement of PID control (or other similar controls) to control the angular position of a DC motor. This network works with supervised learning where data set is presented to train the network before simulation is run to get output results. This ANN controller has two units within it as described below in table 1.

**Table 1**

*Units within ANN Controller*

| ANN Unit | Availability Status of Training Data |
|---|---|
| Controller | Available to train the controller |
| Plant | Available for plant to train the network |

Neural networks in computer world work on the human brain's mechanism of problem solving strategies. In a human brain, several linkages (connections) are provided through networks of axons and synapses to the computing elements called neurons. They communicate to each other in chemical environment where electrical impulses are generated among them to pass information (Kadam et al., 2010; Fukuda & Shibata, 1992). Figure 5 shows the implementation of ANN control for Robot Arm (DC motor position control) through the available model in our MATLAB analysis.

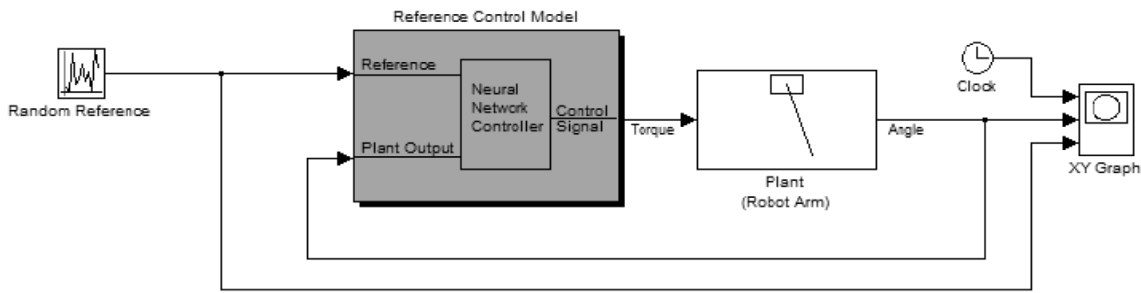

*Figure 5.* ANN control for robot arm (DC motor position control) in MATLAB

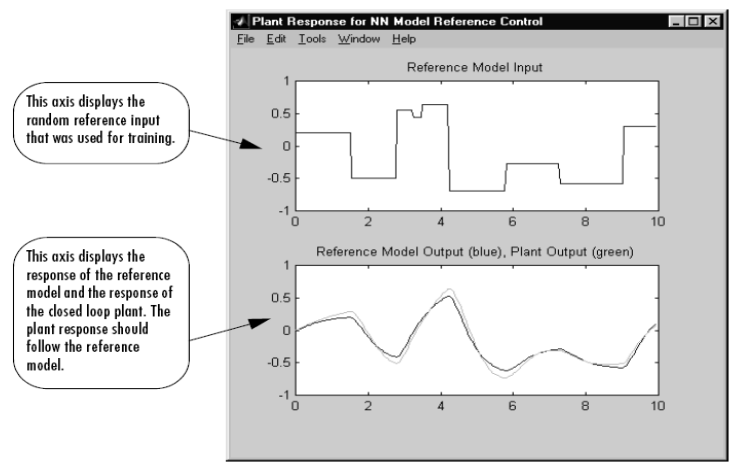

*Figure 6.* Training data generation for ANN controller

Aamir, M.

In figure 6, training data generation for ANN controller in MATLAB is shown. In order to train the controller block of Artificial Neural Network controller, user is free to input the desired values as per the operational requirements before the start of controller's training. In the initial step, the data is generated to train the controller. While data generation process, plant response follows the reference model which is necessary for training's data set to be valid. If the response is not accurate, the data set may be regenerated. If data set is acceptable, the controller may be trained through 'Train Controller' option. The training of Artificial Neural Network controller then starts according to the given parameters. However, it is done after 'Plant Identification' i.e. training the plant unit of ANN controller through the same procedure. The training of ANN controller may take significant amount of time depending upon the given parameters and processing speed. Validation of above training process by the user loads the controller weights into MATLAB's Simulink model.

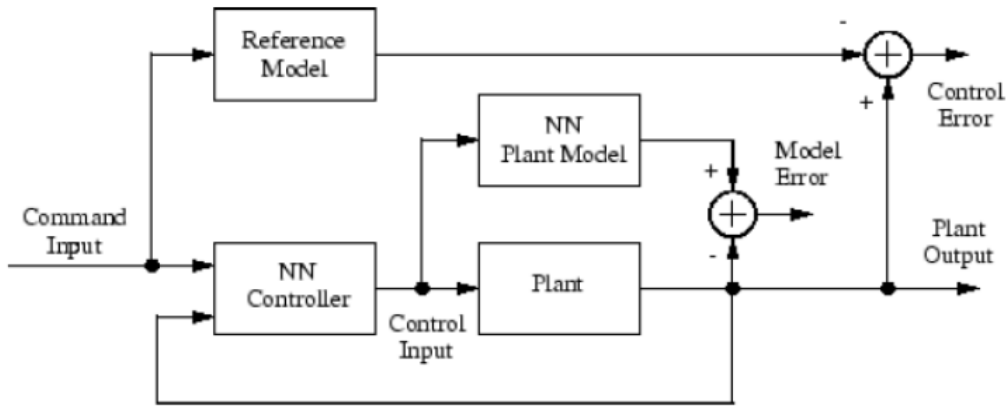

*Figure 7.* Block diagram of the ANN controller

Figure 7 shows the block diagram of ANN controller used in this analysis. NN controller controls the input of plant according to training data. The plant also processes input according to the training data in NN Plant Model. The difference provides the error whereas plant output and control error contribute towards model's output.

Since output of plant is fed back to NN controller's input, it works on back-propagation model of neural networks. Back-propagation neural network is a multilayer feed forward network with back-propagation of an error function (Xu et al., 2012). A simple back-propagation neural network has only three layers i.e. input, output and middle layer as shown below in figure 8.

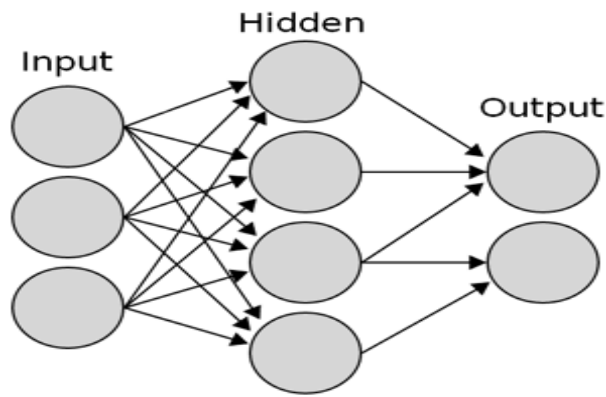

*Figure 8.* Layers of a feed forward neural network

26    *Consortia Academia Publishing*

The input weights are passed on to hidden layer for processing. The hidden layer passes calculated weights to the output layer. The error is presented to input layer through back propagation (feedback) when actual output is different from the desired level. Hence the weights are adjusted to minimize the error through training and learning of the neural network. The process continues until the output is acceptable or pre-configured learning time is achieved (Zhao et al., 2010). In a multi-layer feed forward neural network, more than one hidden layers may be used. Increasing the number of hidden layers provides more accuracy in results but it is harder to implement and increases the cost of system. Therefore, a careful selection of number of hidden layers is required with respect to the cost-benefit analysis.

Multi-layer feed forward neural networks have been used in many real life applications such as image processing and voice recognition. Some major applications in the fields of electrical and electronics engineering utilizing neural networks are robotic applications for position and trajectory controls, speech processing, signal attenuation, pulse width modulation and digital filters etc. (Kadam et al., 2010). The transfer function in feed forward neural networks is usually *sigmoid* function which possesses continuous and nonlinear properties (Fukuda & Shibata, 1992). It is represented by the following equation:

$$f(x) = \frac{1}{1+e^{-ax}} \qquad (4)$$

In recent research attempts, a few neural network models have been proposed for DC motor speed and position controls including motorized robot arm. Zada et al. (2011) proposed neural predictive controller and fuzzy controller to build the block of hybrid control scheme on position control of motorized robot arm. Sabahi (2011) proposed an FEL (feedback error learning) approach to control DC motor's speed through closed loop strategy between PID and ANN controls. In both of the above proposed schemes, it is observed that at least two different controls on each scheme are required to implement their strategies. It can make the models and systems more complicated and costly. On the other hand, neural network's application examined in this paper uses a single ANN control based on backpropagation model of neural networks. Therefore, it can be less complicated and less costly to implement in industrial control applications.

4. **Analysis of Output**

The model shown in figure 5 is simulated in MATLAB's Simulink block.

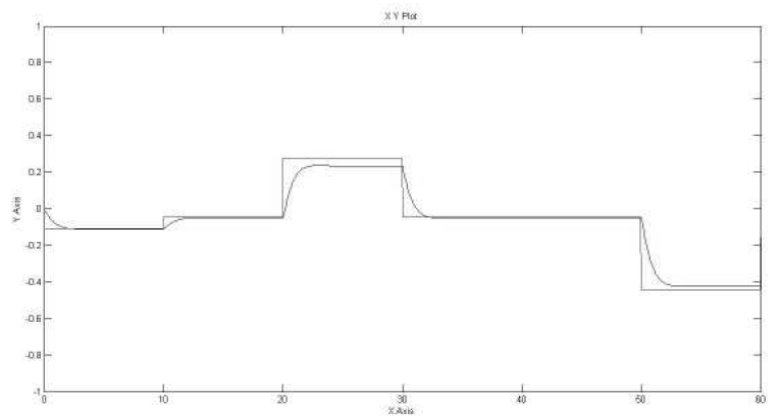

*Figure 9.* Output of ANN control for DC motor position control in robot arm

The simulation result in figure 9 has time value on X Axis and amplitude level on Y Axis. The step inputs are reference signals to the controller whereas the curve represents the response of plant or controlled angular position of robot arm's DC motor to the given step inputs. It is observed that the output of plant follows the input reference signal with acceptable results in terms of time delay factor and other dynamics. The accuracy of ANN control may not be very high, but still comparable to PID control with acceptable results. The behavior of ANN control to unknown dynamics is relatively better than PID control. Although PID control also takes care of unknown dynamics but ANN control learns from experience as Artificial Neural Networks are trained through data set in supervised learning. Therefore, ANN control is more responsive to unknown dynamics of the system. It makes it more suitable for industrial control applications as an industrial control system also has uncertainties and time-varying effects.

We analyze the output of the system by varying the network size of ANN controller. The analysis is done in terms of Mean Square Error (MSE) representing the difference between desired and actual output levels. In ideal conditions, the MSE approaches to zero. We use three different network sizes (number of input layer neurons in correspondence to the output layer neurons). The input neurons are selected as 5, 10 and 15 for single output neuron. Number of epochs is fixed at 500. The results are presented in table 2.

**Table 2**

*ANN Controller test results with different network sizes*

| Network Size | No. of Epochs | Mean Square Error (MSE) |
|---|---|---|
| 5-1 | 500 | 4.34 |
| 10-1 | 500 | 0.21 |
| 15-1 | 500 | 0.0015 |

The above test results show that increasing the number of input neurons or network size remarkably improves the system's performance in terms of Mean Square Error (MSE). A lower value of MSE is always desired for an effective control system. Hence, ANN based DC motor position control for plant units like a robot arm is suitable to implement and cope with uncertain system dynamics due to its well known training and learning capabilities.

### 5. Conclusion and Future Directions

In this paper, we discussed the replacement of PID controller in industrial application of DC motor position control for robot arm with ANN controller based on back-propagation model of neural networks. The simulation results in MATLAB showed that the output of plant in the examined ANN control follows the input reference signal with acceptable results in terms of time delay factor and system dynamics. We also analyzed that increasing the network size further improves the overall system's performance in terms of Mean Square Error (MSE). Since ANN control learns from experience as it is trained through data set in supervised learning, ANN control is more responsive than PID to unknown dynamics of the system which makes it even more suitable for industrial control applications having uncertainties and unknown dynamics due to environmental noise. This research is focused on evaluating the feed forward neural network with sigmoid transfer function and back-propagation of error. In future, we may include other processes and plant units of the process industry and simulate the results for observing the applicability of back-propagation ANN control on them. There may be more logics and enhanced training procedures for ANN controller to gain more precise results in reference control applications.